%% file: bluespec3.tex
\documentclass[12pt]{article}
\usepackage{epsfig,amssymb,bm,graphicx,amsmath,float}
\usepackage[normalem]{ulem}
\usepackage{amsmath}
\usepackage{graphicx}

\newcommand\pubnumber{}
\newcommand\pubdate{\today}

\def\napoli{Department of Physics\\
University of Notre Dame, Notre Dame, IN, 46556, USA}
\def\support{\footnote{jbraman2@nd.edu}}

\def\Title#1{\begin{center} {\Large #1 } \end{center}}
\def\Author#1{\begin{center}{ \sc #1} \end{center}}
\def\Address#1{\begin{center}{ \it #1} \end{center}}

\newcommand\pubblock{\rightline{\begin{tabular}{l} \pubnumber\\
         \pubdate  \end{tabular}}}
\newenvironment{Abstract}{\begin{quotation}  }{\end{quotation}}
\newenvironment{Presented}{\begin{quotation} \begin{center} 
             PRESENTED AT\end{center}\bigskip 
      \begin{center}\begin{large}}{\end{large}\end{center} \end{quotation}}
\def\Acknowledgements{\bigskip  \bigskip \begin{center} \begin{large}
             \bf ACKNOWLEDGEMENTS \end{large}\end{center}}

\input econfmacros.tex

\begin{document}
\begin{titlepage}
\pubblock

\vfill
\Title{Beyond the Hubble scale: Super cosmic variance and nongaussianity as a portal to the superhorizon}
\vfill
\Author{Joseph Bramante\support}
\Address{\napoli}
\vfill
\begin{Abstract}If cosmological perturbations in our Hubble sized volume are nongaussian, then they will be coupled to any larger perturbation modes outside our Hubble volume. This has important consequences for modeling inflation: the scalar power spectrum and spectral index measured in our Hubble volume would depend on an adjacent background of super Hubble perturbations. In other words, a detection of nongaussianity implies a possible portal to the superhorizon. By the same token, ruling out nongaussianity would rule out the possibility that the power spectrum's size and running are accidents of super cosmic variance. In this note, we provide a compact derivation of super cosmic variance, survey recent results, and show how experimental limits on nongaussianity help to rule it out. 
\end{Abstract}
\vfill
\begin{Presented}
 The 10th International Symposium 
on Cosmology and Particle Astrophysics (CosPA2013)\\
Honolulu, HI, USA,  November 12--15, 2013
\end{Presented}
\vfill
\end{titlepage}
\def\thefootnote{\fnsymbol{footnote}}
\setcounter{footnote}{0}
\section{Super cosmic variance}
The impact of super cosmic variance on cosmological observation has been demonstrated in recent work \cite{Nelson:2012sb,Nurmi:2013xv,LoVerde:2013xka,Bramante:2013moa,LoVerde:2013dgp}, which has shown that measurements of the bispectrum, trispectrum, power spectrum, and the runnings of the same \cite{Bramante:2013moa} will vary between different Hubble scale volumes in a nongaussian universe with super Hubble extent. 

The mechanism responsible for this super cosmic variance can be explained in a few paragraphs. The cosmological scalar curvature perturbation $\zeta$ parameterizes spatially-dependent fluctuations of the primordial scale factor $a$. 
\begin{align}
a(x) = \bar{a} e^{\zeta(x)}
\end{align}
Here $\bar{a}$ is the average value of the scale factor, such that the average of scalar curvature perturbations over the volume will be zero, $\bar{\zeta} = 0$. During a period of cosmological inflation, the comoving scale $(aH)^{-1}$ rapidly shrinks before expanding during the epoch of reheating. But it is the dynamics of reheating which determine the extent to which the comoving scale $(aH)^{-1}$ will grow after inflation. Therefore, it is plausible that inflation occurred over scales larger than those visible in our Hubble horizon. The scalar curvature perturbation, generated at some time during inflation and reheating, is customarily expressed as a sum over a gaussian ($\zeta_G$) and additional nongaussian terms,
\begin{align}
\zeta (\vec{k}) = \zeta_G (\vec{k}) + \frac{3}{5} f_{\rm{NL}} (k) \int \frac{d^3\vec{p}}{(2 \pi)^3} \zeta_G (\vec{p}) \zeta_G (\vec{k}-\vec{p}) + \cdots, \label{zeta}
\end{align}
Here we have cast $\zeta$ in momentum space both for convenience and to make our point. If we remove all nongaussian terms in Eq. \ref{zeta}, that is set $f_{\rm{NL}}$ and all subsequent nongaussian coefficients to zero, then it is obvious from momentum conservation that the resulting scalar curvature perturbation will only correlate perturbations of the same wavelength. That is, if $\left\langle \zeta (\vec{k}) \zeta (\vec{k'}) \right\rangle = \left\langle \zeta_G (\vec{k}) \zeta_G (\vec{k'}) \right\rangle \neq 0$, then $\vec{k} = -\vec{k'}$ -- these are the only two momenta available if the curvature perturbation is completely gaussian. 

On the other hand, nongaussian curvature perturbations with $f_{\rm{NL}},g_{\rm{NL}}\ldots \neq 0$ allow for perturbation modes of different wavelengths to be correlated. Indeed, the so-called ``local" nongaussian term $f_{\rm{NL}}$ in Eq. \ref{zeta} attains its largest values when $|\vec{k}_1| \ll |\vec{k}_2|,|\vec{k}_3| $ for the bispectrum $\left\langle \zeta (\vec{k}_1) \zeta (\vec{k}_2) \zeta (\vec{k}_3)  \right\rangle$. But if perturbation modes of different wavelengths are correlated, then by definition there must be regions of short wavelength perturbations which are anomalously high or low, collocated with long wavelength perturbations which are anomalously high or low -- for if not, then long and short wavelength perturbations are simply uncorrelated, and the statistics are gaussian.\footnote{An exception would be purely equilateral nongaussianity, where longer perturbations are uncorrelated with shorter perturbations. In that case, there is nongaussianity without large wavelength perturbations necessarily correlating with small. Indeed, one should not expect purely equilateral nongaussianity to induce super cosmic variance, as discussed in \cite{Bramante:2013moa}.}

The program of super cosmic variance, then, is to understand how anomalously high or low sums of super Hubble sized perturbations correlate with measurements of smaller perturbations inside the Hubble horizon. When we say \emph{anomalously} high or low, it is important to emphasize that observationally significant overdensities and underdensities of superhorizon modes should be expected in a universe sufficiently larger than our Hubble patch, depending on the model sourcing cosmological perturbations. To illustrate how super cosmic variance depends on the size of the super Hubble universe and the sum over super Hubble scale perturbations located at our Hubble patch, we calculate the power spectrum using Eq. $\ref{zeta}$. The power spectrum is typically defined $(2 \pi)^3 \delta^3(\vec{k} + \vec{k'})P(k)\equiv \left\langle \zeta (\vec{k}) \zeta (\vec{k'}) \right\rangle $, so the power spectrum for Eq. \ref{zeta} is proportional to
\begin{align}
P(k) \propto \left\langle \zeta_G (\vec{k}) \zeta_G (\vec{k'}) \right\rangle + \frac{6}{5}  f_{\rm{NL}} (k)\left\langle \int \frac{d^3\vec{p}}{(2 \pi)^3} \zeta_G (\vec{p}) \zeta_G (\vec{k}-\vec{p}) \zeta_G (\vec{k'}) \right\rangle + \cdots, \label{power}
\end{align}
where for the moment we discard the term proportional to $f_{\rm{NL}}^2 (k)$ as small. Readers may note that the last term of Eq. \ref{power} is an expectation of an odd number of gaussian fields and wonder why this term has not also been discarded. For subhorizon modes, this term is indeed zero. However, let us now consider the portion of the integral in Eq. \ref{power} that runs over super Hubble wavemodes and likewise assume that $|\vec{k}| \gg |\vec{p}|$ so that we may take $\zeta_G (\vec{k}-\vec{p}) \simeq \zeta_G (\vec{k})$. In that case the power spectrum is given by 
\begin{align}
P(k) \propto \left\langle \zeta_G (\vec{k}) \zeta_G (\vec{k'}) \right\rangle \left(1 + \frac{12}{5}  f_{\rm{NL}} (k)  \left\langle \int\limits_{|\vec{p}| < H_0} \frac{d^3\vec{p}}{(2 \pi)^3}  \zeta_G (\vec{p})\right\rangle \right), \label{power2}
\end{align}
where here we have made the choice that $\vec{k}$ is a Hubble scale wavemode and the integral over $\vec{p}$ will be over super Hubble wavemodes. 

Now from the perspective of a ``super observer" who could measure all super Hubble scale modes, the term $ \left\langle \int\limits_{|\vec{p}| < H_0} \frac{d^3\vec{p}}{(2 \pi)^3}  \zeta_G (\vec{p})\right\rangle$ is indeed null. As previously mentioned, this framework is constructed so that the total spatial average of $\zeta_G$ is zero. However, from the perspective of an observer with access to only one patch of Hubble length, this term will be the sum over superhorizon perturbations $\zeta_G$, evaluated at the location of a particular Hubble patch. 

At this point it cannot be over-emphasized that assuming the last piece of Eq. \ref{power2} is equal to zero is the same as assuming that all super Hubble perturbations evaluated for a particular Hubble patch sum exactly to zero. While this will be nearly the case for some Hubble sized patches, it is not generally true. Furthermore, the chance that all super Hubble perturbations cancel will shrink for larger super Hubble volumes -- the variance of the sum over super Hubble modes will increase.

It is useful to transform this integral over superhorizon perturbations back to position space and define a term quantifying the bias of super Hubble modes. Using the conventions of \cite{Bramante:2013moa}, we define the sum over super Hubble perturbations in our Hubble patch as $\zeta_{Gl}$ and the bias as $B= \zeta_{Gl}/ \left\langle \zeta_{Gl}^2 \right\rangle^{1/2}$. We note in passing that the value of $ \left\langle \zeta_{Gl}^2 \right\rangle$ can be determined from the power spectrum outside the Hubble horizon. The observed power spectrum, $P^{\rm{obs}}(k)$ is then given by\footnote{In order to simplify and shorten this exposition we have left out that the observed scalar curvature perturbation will technically equal $\zeta^{\rm{obs}} = \zeta_s/(1+\zeta_l)$, where $\zeta_s,\zeta_l$, are curvature perturbations sequestered to wavemodes respectively smaller and larger than the Hubble scale. This difference does not affect the results of this note so long as $\zeta_l \ll 1$, which we assume throughout this treatment. For a full exposition see the introduction of \cite{Bramante:2013moa}.}
\begin{align}
P^{\rm{obs}}(k) = P(k) \left(1 + \frac{12}{5}  f_{\rm{NL}} (k)  B \left\langle \zeta_{Gl}^2 \right\rangle^{1/2}  \right), \label{power3}
\end{align}
where now $P(k)$ is the power spectrum that a super observer with access to all superhorizon modes would measure at scale $k$. The super observer perspective is an important notion from a model building standpoint: any model predicting nonequilateral nongaussianity and an inflating volume larger than the Hubble scale will additionally imply that our cosmological observables vary from one Hubble patch to another by an amount which depends on the size of $\left\langle \zeta_{Gl}^2 \right\rangle^{1/2}$.

\section{Survey of super cosmic variance}
Calculations of super cosmic variance have led to striking conclusions about the relationship between our Hubble scale and a wider universe. Superhorizon perturbations may lead local observations to drastically depart from superhorizon statistics. For example \cite{Nelson:2012sb} demonstrated that our Hubble volume may sit in a super Hubble universe with exclusively nongaussian statistics. This is worth spelling out: if the scalar curvature perturbation is exactly $\zeta = f_{\rm NL} \zeta_G^2$, then given a sufficiently large super Hubble universe, an observer in a Hubble patch can measure perturbations which are gaussian $\zeta^{\rm{obs}} = \zeta_G$. 

Work in \cite{Nurmi:2013xv} and \cite{LoVerde:2013xka} quantified how subhorizon power spectra and nongaussianities can vary with as few as ten extra e-foldings of superhorizon perturbations. \cite{LoVerde:2013xka} especially showed how these affects extend to large $n$-point functions. While the recent spate of papers have thoroughly explored super cosmic variance, it is interesting that the first examples of Hubble patches with varying statistics appeared in curvaton models \cite{Linde:2005yw,Demozzi:2010aj}. 

Although this note has mostly addressed super cosmic variance from a statistical perspective, there is an equivalent perspective which says that inflating fields will arrive at different field values in different Hubble patches as they roll down their potentials. This equivalence of super cosmic variance (1) as a shift in field values and (2) as a shift in small scale statistics from superhorizon mode coupling has been explicitly calculated for curvaton models in \cite{LoVerde:2013dgp}.

Super cosmic variance of the spectral index and nongaussian running were investigated in \cite{Bramante:2013moa}. Quite remarkably, it was found that with a modest bias from superhorizon perturbations and with running local nongaussianity, the scalar spectral index may actually be blue over Hubble scales from the perspective of a super observer. Superhorizon scalar modes were also found to shift subhorizon tensor perturbations, a topic further developed in \cite{Brahma:2013rua}. Finally, \cite{Bramante:2013moa} addressed how one might close the nongaussian portal to the superhorizon: tighter limits on nongaussianity correspond to tighter limits on super cosmic variance, on which we elaborate in the remainder of this note.

\section{Limits on super cosmic variance from limits on nongaussianity}
\begin{figure}[H]
\caption{Limits on superhorizon values of $f_{NL}(k)$ and $\left\langle \zeta_{Gl}^2 \right\rangle$ are displayed, under the assumption that our Hubble volume sits on either a $2 \sigma$ overdense or underdense sum of superhorizon perturbations. Bounds on superhorizon parameters are shown assuming $f_{\rm{NL}}^{\rm{obs}}(k)$ is positive and either less than ten or one. Nonperturbative regions of parameter space are blocked out where the the scalar curvature perturbation is no longer a small perturbation of the scale factor $a$, ($|\zeta| > 0.1$).}
\center
\label{fig:limits}
\begin{tabular}{cc}
\includegraphics[scale=.8]{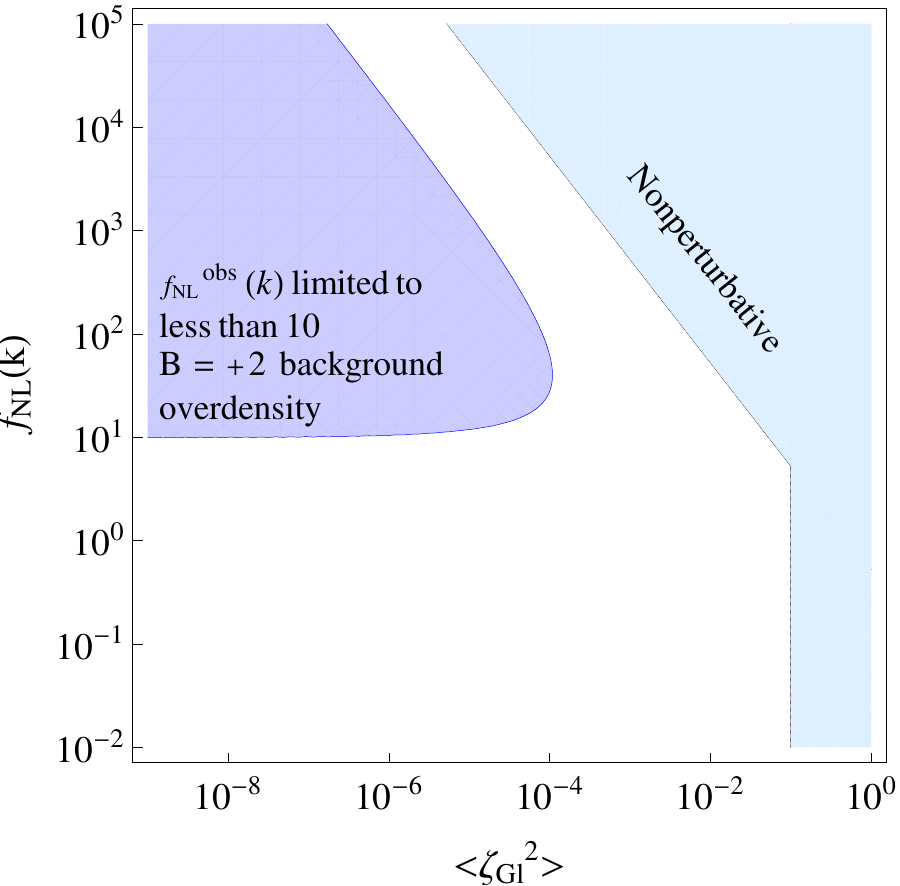} & \includegraphics[scale=.8]{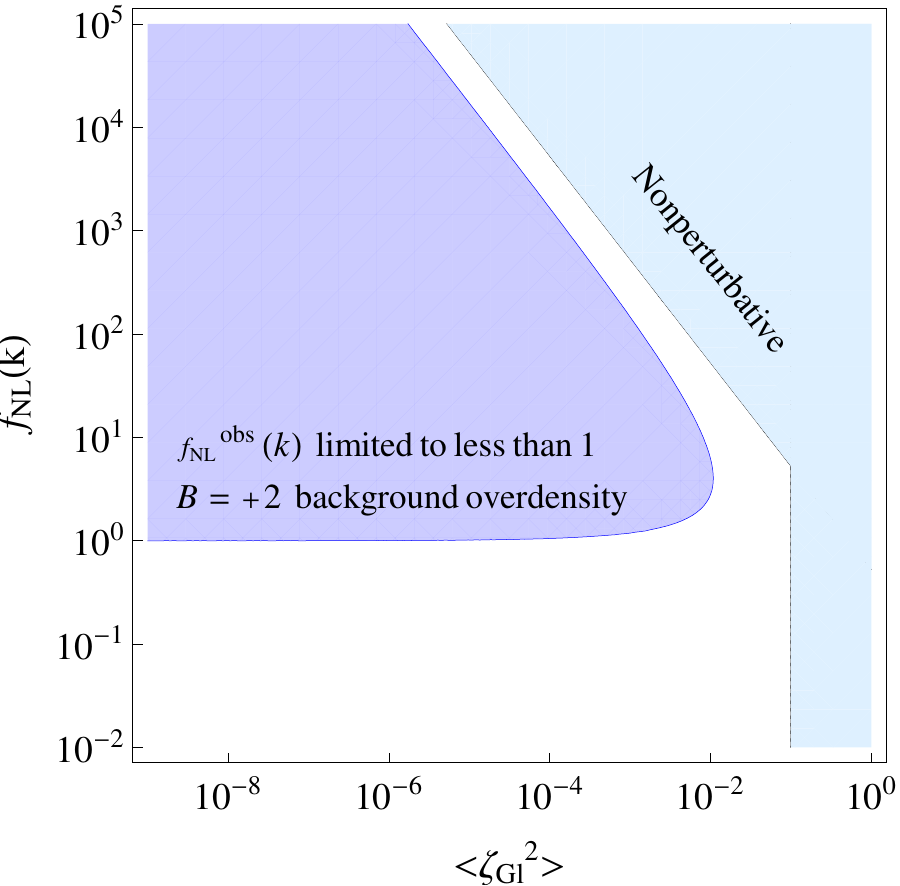} \\
\includegraphics[scale=.8]{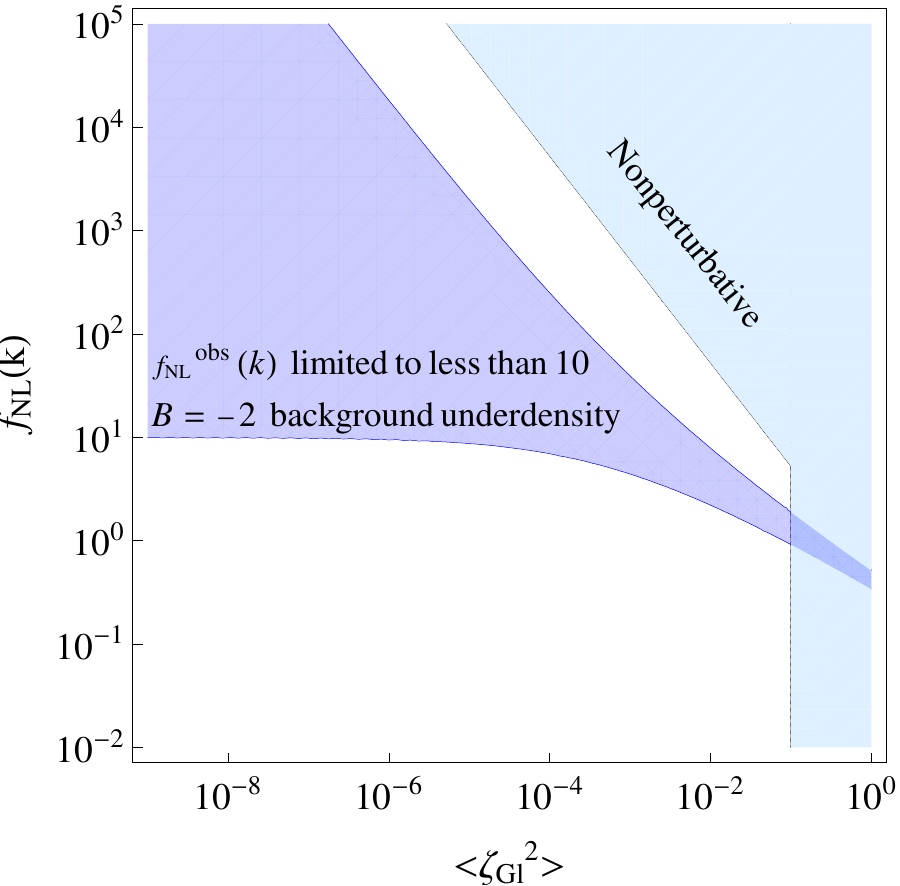} & \includegraphics[scale=.8]{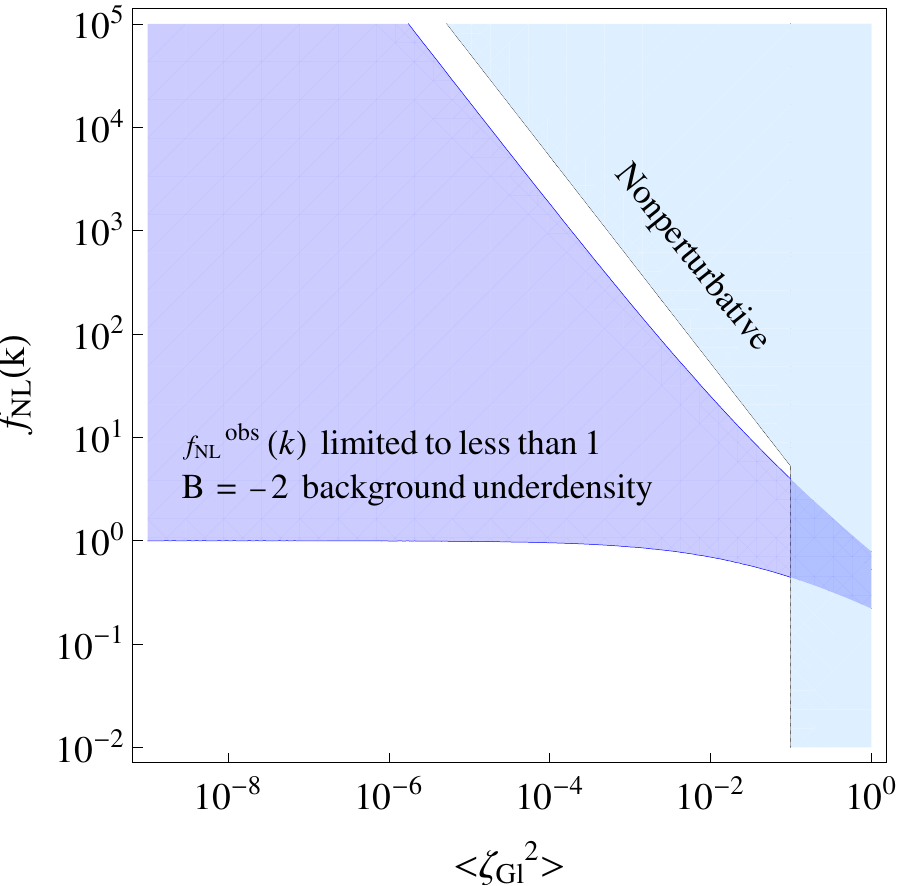}
\end{tabular}

\end{figure}

Recent work on super cosmic variance has made the task of matching cosmological observation to models of inflation more interesting -- superhorizon structure makes heretofore excluded nongaussian models viable. It has also shown how to shut the portal on super cosmic bias: tighten bounds on mode correlation inside the Hubble horizon. Assuming the curvature perturbation has only a linear and quadratic term, the observed value of $f_{\rm{NL}}^{\rm{obs}}(k)$ can be related to the value of $f_{\rm{NL}}(k)$ as seen by a super observer,\footnote{See \cite{Nurmi:2013xv}, \cite{LoVerde:2013xka}, and \cite{Bramante:2013moa} for a complete treatment}
\begin{align}
f_{\rm{NL}}^{\rm{obs}}(k) = \frac{f_{\rm{NL}}(k)}{\left(1+\frac{6}{5} f_{\rm{NL}}(k) B \left\langle \zeta_{Gl}^2 \right\rangle^{1/2}  \right)^2},
\end{align}
where again we identify $\left\langle \zeta_{Gl}^2 \right\rangle$ as the background perturbation variance determined by the power spectrum outside the Hubble horizon, while $B$ is a measure of the biasing of a given Hubble patch of the universe. In Figure \ref{fig:limits} we show that even assuming our Hubble patch sits on a substantial over or underdensity, i.e. $B=+2$ or $B=-2$, a wide swath of superhorizon parameter space can be ruled out by bounding the observed local nongaussianity, $f_{\rm{NL}}^{\rm{obs}}(k)$. The crucial point here is that super cosmic variance will depend on the \emph{subhorizon} value of nongaussianity, and this can be limited.

In this note we have only considered models which have nongaussianity of a quadratic, local type, $f_{\rm{NL}}^{\rm{local}}$. We should also point out that these formulae rely on the assumption that $|\vec{p}| \ll |\vec{k}|$, or in other words we assume that most of the biasing occurs from wavemodes much longer than Hubble scale modes, leaving the affect of large near horizon perturbations an open question. Note also in Figure \ref{fig:limits} that while we block out the nonperturbative regions of superhorizon parameter space, developing a method to relate super cosmic bias to subhorizon statistics in the case that $\left\langle \zeta_{Gl}^2 \right\rangle$ grows to nonperturbative values ($\left\langle \zeta_{Gl}^2 \right\rangle \gtrsim 1$) is an interesting topic for future research.

\Acknowledgements
I would like to thank Jason Kumar, Elliot Nelson, and Sarah Shandera for collaboration.

\end{document}

%% file: econfmacros.tex
\def\beq{\begin{equation}}
\def\eeq#1{\label{#1}\end{equation}}
\def\eeqn{\end{equation}}

\def\beqa{\begin{eqnarray}}
\def\eeqa#1{\label{#1}\end{eqnarray}}
\def\eeqan{\end{eqnarray}}

\let\bar=\overbar

\def\Dslash{\not{\hbox{\kern-4pt $D$}}}
\def\dslash{\not{\hbox{\kern-2pt $\del$}}}

\def\msb{{\bar{\ssstyle M \kern -1pt S}}}

